\documentclass{iau} 
\usepackage{graphicx}

\title[JD 11.~~Reconciling SGREs and SEPs] 
{The Relationship between Long-Duration Gamma-Ray Flares and Solar Cosmic Rays}

\author[Hugh S. Hudson]   
{Hugh S. Hudson$^{1,2}$}

\affiliation{$^1$Space Sciences Laboratory, University of California, Berkeley CA USA 94720
 \\ email: {\tt hhudson@ssl.berkeley.edu} \\[\affilskip]
$^2$SUPA School of Physics and Astronomy, University of Glasgow, UK}

\pubyear{2017}
\volume{335}  
\jname{Space Weather of the Heliosphere: Processes and Forecasts}
\editors{Claire Foullon \& Olga Malandraki}
\begin{document}

\maketitle

\begin{abstract}
A characteristic pattern of solar hard X-ray emission, first identified in SOL1969-03-30 by Frost \& Dennis (1971), turns out to have a close association with the prolonged high-energy gamma-ray emission 
originally observed by Forrest \textit{et al.} (1985).
This identification has become clear via the observations of long-duration $\gamma$-ray flares by the  \textit{Fermi}/LAT experiment, for example in the event SOL2014-09-01. 
 \nocite{1971ApJ...165..655F}
 \nocite{1985ICRC....4..146F}
The distinctive features of these events include flat hard X-ray spectra extending well above 100 keV, a characteristic pattern of time development, low-frequency gyrosynchrotron peaks, CME association, and gamma-rays identifiable with pion decay originating in GeV ions. 
The identification of these events with otherwise known solar structures nevertheless remains elusive, in spite of the wealth of EUV imagery available from SDO/AIA. 
The quandary is that these events have a clear association with SEPs in the high corona, and yet the gamma-ray production implicates the photosphere itself, despite the strong mirror force that should focus the particles \textit{away} from the Sun
We discuss the morphology of these phenomena and propose a solution to this problem.
\end{abstract}

\firstsection 
\section{Introduction}

Flares and CMEs involve substantial particle acceleration, and we can see the consequences of this in several forms. 
Radio-astronomical techniques allow us to remote-sense some populations of energetic electrons in the corona, but such information at high energies often involves the transport of particles to regions of high density where emission can form.
For electrons, the basic diagnostic radiation is bremsstrahlung, while for accelerated ions it is a variety of nuclear processes, such as the inelastic scattering of high-energy primary particles on ambient nuclei.

The processes involved with particle acceleration in the corona remain ill-understood, mainly because of the weak diagnostic power of the direct emissions (e.g., Krucker et al. 2010) and the difficulty of interpretation of the radio signatures (e.g. Kundu, 1976).
\nocite{1978ApJ...224..235H}
\nocite{2015ApJ...805L..15P}
\nocite{2010ApJ...714.1108K}
\nocite{1965sra..book.....K}
Both electrons and ions contribute to solar particle events detectable in the heliosphere, but the identification of these emissions with specific structures in the corona, as identified for example by soft X-ray or EUV imaging, remains largely unclear.
In the heliosphere, for example, we have a clear association of particle acceleration with interplanetary shock waves, but we have little information about how these shocks form, and how they attain their energy. 

Recently these problems have become especially prominent because of the detection of many long-duration events by the \textit{Fermi}/LAT experiment, at energies above 20~MeV.
Share \textit{et al.} (2017) coined the term ``sustained gamma-ray events'' (SGREs) to imply a source population possibly distinct from that of the normal flare $\gamma$ rays.
Their detected photon spectra can extend above 1~GeV and display the characteristics of pion decay, again implying GeV particle energies.
Table~\ref{tab:prop} summarizes the properties of such events, based on the ``discovery'' events in hard X-rays (Frost \& Dennis, 1971) and $\gamma$-rays (Forrest \textit{et al.},1985).
\nocite{2017arXiv171101511S}
\nocite{2017spw.book.....K}
In principle any sustained $\gamma$-ray emission could come either from continued particle acceleration
or from the gradual interaction of trapped particles.
Mandzhavidze \& Ramaty (1992) favored the latter, as further discussed here, and noted a requirement for the mirror ratio to exceed 10 and the ambient density to lie below $5 \times 10^{11}$~cm$^{-3}$, ``quite reasonable'' conditions.
Klein \textit{et al.} (2017) also discuss the mirror force in this context, though Plotnikov \textit{et al.} (2017) do not.
\nocite{2017arXiv170307563P}

\begin{table}
  \begin{center}
  \caption{Properties of coronal high-energy radiations.}
  \label{tab:prop}
  \begin{tabular}{|l|l|}\hline 
 {\bf Hard X-rays} & {\bf High-energy $\gamma$-rays} \\
 \hline
 SOL1969-03-30$^1$  & SOL1982-06-03$^2$ \\
 \hline
 Coronal origin (by occultation)$^3$ & Coronal origin (by inference)$^4$ \\
 Hard spectrum, $J \sim (h\nu)^{-2}$ & Pion decay signature \\
 Duration tens of minutes & Duration up to many hours \\
 Association with type II/IV burst$^5$ & Association with type II/IV burst \\
 Drifting cm-wave source$^6$ & Neutrons \\
 Association with SEPs & Association with SEPs \\
 Large angular scale$^7$ & Unknown structure \\
 \hline
  \end{tabular}
 \end{center}
\vspace{1mm}
 {\it Notes:}
  $^1$Frost \& Dennis, 1971. 
  $^2$Forrest \textit{et al.}, 1985.
  $^3$Hudson, 1978.
  $^4$Chupp \textit{et al.}, 1985.
  $^5$Smerd, 1970.
  $^6${\' E}nom{\' e} \& Tanaka, 1971.
  $^7$Krucker \& Lin, 2008.
\end{table}
\nocite{1971IAUS...43..413E}
\nocite{2008ApJ...673.1181K}
\nocite{1970PASAu...1..305S}
\nocite{1985ICRC....4..126C}

\begin{figure}[h]
\begin{center}
 \includegraphics[width=0.49\textwidth]{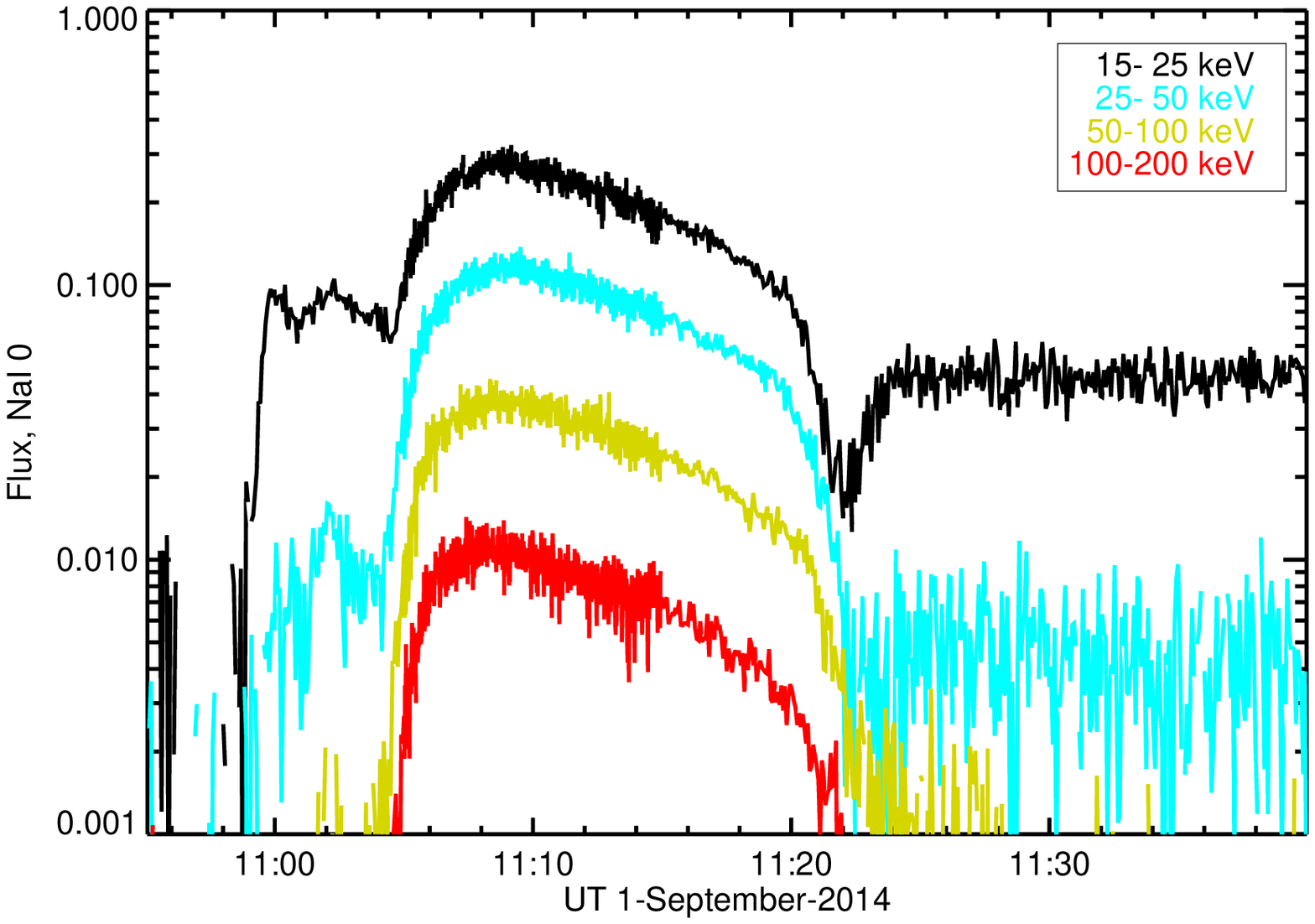} 
 \includegraphics[width=0.49\textwidth]{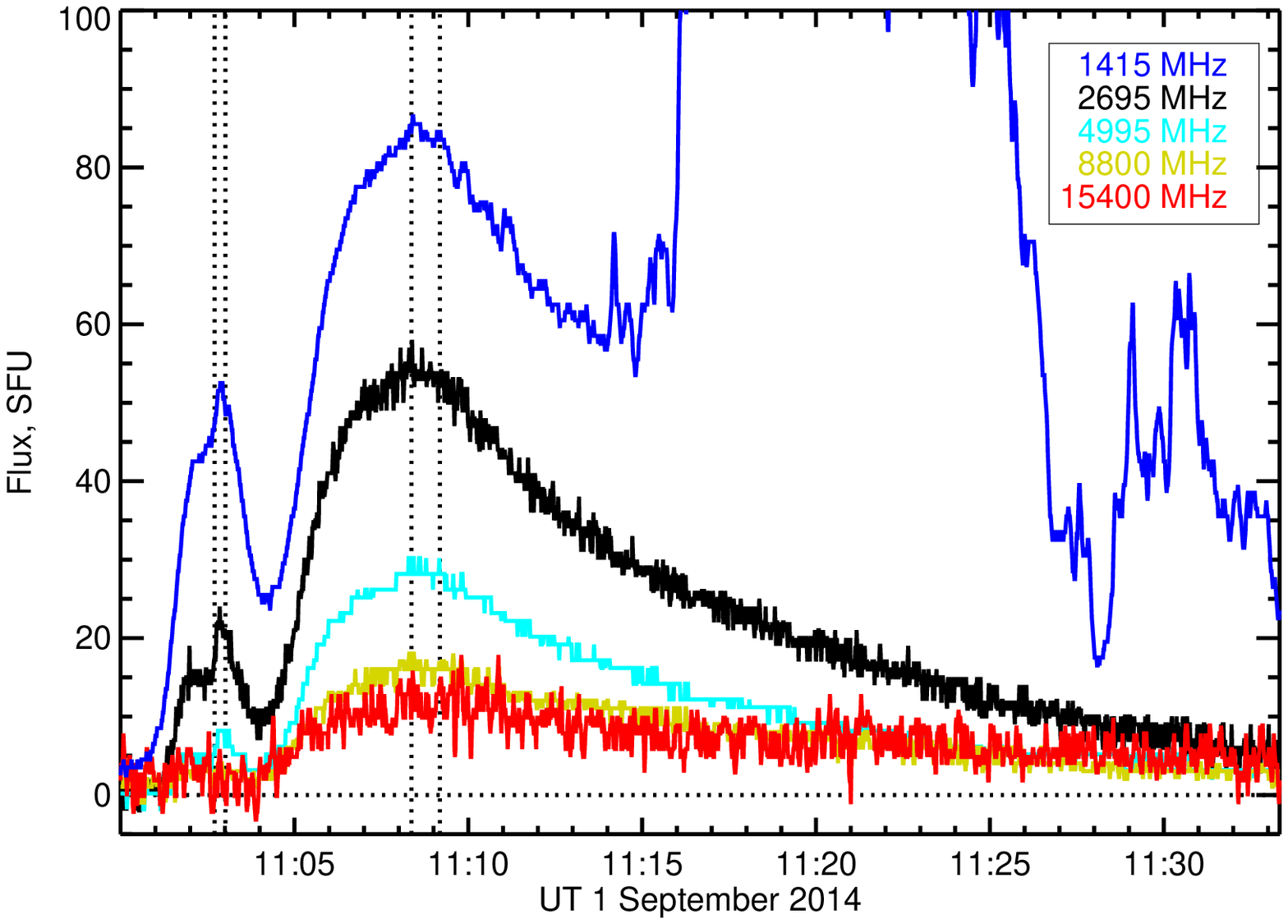} 
 \caption{SOL2014-09-01, observed in hard X-rays by \textit{Fermi}/GBM (log scale, left) and by the San Vito radio observatory (linear scale, right). 
 The light curves match extremely well, except at longer radio wavelengths at which a distinct spectral component appears.
 }
   \label{fig:SOL2014-09-01}
\end{center}
\end{figure}

Recently a ``Rosetta Stone'' event appeared (Pesce-Rollins \textit{et al.} 2015; Ackermann \textit{et al.} 2017)
\nocite{2015ICRC...34..128P,Ackermann_2017A}, 
in which both a Frost-Dennis hard X-ray source and sustained high-energy $\gamma$-ray emission occurred together, almost coincident in time but with the $\gamma$ rays remaining detectable for almost two hours.
This event, SOL2014-09-01, as shown in Figure~\ref{fig:SOL2014-09-01}, matches the HXR/radio properties in Table~\ref{tab:prop} quite well.
As inferred previously for Frost-Dennis events, the extreme flatness of the HXR spectrum and its association with a matching non-thermal microwave spectrum, with low turnover frequency, strongly implicates relativistic electrons in the corona.
In this event \textit{Fermi}/LAT also detected high-energy $\gamma$-radiation.

\section{The mirror force and particle numbers}

The association of Frost-Dennis/SGRE events with SEPs naturally suggests the idea that the SEPs themselves return to the Sun and interact to produce the HXR and $\gamma$-ray emission
(Pesce-Rollins \textit{et al.} 2015; De Nolfo  \textit{et al.} 2016, Share \textit{et al.} 2017, Ackermann \textit{et al.} 2017).
\nocite{Ackermann_2017A}
\nocite{2016AGUFMSH51E2616D}
This interpretation, however, immediately seems counterintuitive.
The conventional explanation of SEPs describes them as the result of diffusive shock acceleration in a bow wave driven by the CME.
Such a shock would form in the middle corona, where a relatively weak magnetic field might favor shock formation.
This very weakness of the field implies a strong mirror force that would tend to ``focus'' the particles away from the Sun, especially on open field structures.
The probability for an accelerated particle to fall in the loss cone defined by collisions at photospheric height would be given by a probability $P  = 1 - \sqrt{1 - B/B_0}$, where $B$ is the ambient field intensity and $B_0$ the intensity at the precipitation footpoints.
Note that a similar mirror force would apply to a diffusive SEP propagation, but that in any case we really do not have any quantitative knowledge of the actual geometry of the coronal magnetic field.
For the SolarSoft PFSS field extrapolation (Schrijver \& DeRosa 2003) at the time of SOL2014-09-01, we find that $\bar{P} = \Delta \Omega/4/\pi \approx 0.003$ for closed field, based on the mean field strengths at source surface at 2.5~R$_\odot$ and the photosphere.
As an independent crude estimate, a dipole field originating 0.1~R$_\odot$ below the photosphere would have a mirror ratio $P = 0.0008$.

A more basic question about the association between SEPs and SGREs comes from the total numbers of particles needed.
\nocite{2005ESASP.592...67M}
Mewaldt \textit{et al.} (2005) suggest particle fluences of order 0.1-10~protons~(cm$^2$ sr MeV)$^{-1}$ at 500~MeV, for a set of five well-observed events.
This estimate includes rough approximations for multiple counting and for longitude dependence.
Converted to energy fluxes, assuming a 10\% propagation cone, these fluences imply total particle energies of $2 \times 10^{24-26}$~erg above 500~MeV.
Ackermann \textit{et al.} estimate at least $2.5 \times 10^{24} - 10^{27}$~erg for the energy required to sustain the occulted SGRE events they study.
So, even not considering the mirror force, there is a substantial number problem: more particles appear to be required to produce the $\gamma$-rays than even the large energy fluxes in the interplanetary SEPs.

\section{A ``Lasso'' Scenario}

Based on the above, the suggestion that SEP particles can somehow produce the SGRE emissions seems highly unlikely.
I propose a ``Lasso'' scenario to enable this relationship, in which distended but closed magnetic fields would form a noose that could capture some of the SEPs, not allowing their escape into the heliosphere.
In this picture the particle accelerator deposits particles on both open and closed field structures.
The closed-field domain might have a large spatial scale, perhaps extending to several R$_\odot$ in height.
This then bodily retracts, as a part of the recovery of the corona to the CME eruption, and thereby 
transporting the non-thermal particles by advection into denser regions. 
In this process one would also expect the betatron process and first-order Fermi acceleration to increase the particle energies and effectively to multiply their number relative to the SEPs themselves (the Compton-Getting effect).
This idea in a sense modernizes the original suggestion by Elliot (1969) for particle storage in the low-beta corona.
\nocite{1969sfsr.conf..356E}
The Lasso idea just imagines that closed field can ensnare some of the Elliot particles and bring them down towards the Sun, where increased density and/or actual precipitation can produce the nuclear interactions needed.
Post-CME inward flows on large spatial scales, as observable by LASCO, are not unusual (Sheeley \textit{et al.} 2004, 2014), as required by the heliospheric regulation of open solar flux.
\nocite{2004ApJ...616.1224S}
\nocite{2014ApJ...797...10S}
Indeed the readily-available LASCO movies show hints of an inflow at the N flank of the CME in the SOL2014-09-01 event, in a manner consistent with the Lasso picture.

Do we actually require the lasso action (retraction of large-scale magnetic flux bundles) to find sufficient energetic particles? A static field volume at a sufficiently low density might in principle contain 500~MeV protons long enough to generate pions and $\gamma$-ray continuum \textit{in situ} via trapping in turbulence.
The relativistic Bethe-Bloch cross-section for collisional losses is about $3 \times 10^{-23}$~cm$^2$ (Olive \textit{et al.} 2014), which for density 10$^8$~cm$^{-3}$ translates into a time scale of five hours;
for these parameters the Alfv{\' e}n speed is about 3000~km/s.
\nocite{2014ChPhC..38i0001O}
Thus a ``coronal thick target'' scenario could provide an alternative to the Lasso picture suggested here.
 
\section{Conclusions}

The events described here (the Frost-Dennis and SGRE phenomena) appear to form a distinctive pattern, and one that probably happens fairly commonly during active periods.
The feebleness of the diagnostic radiations in hard X-rays and $\gamma$-rays makes their detection difficult, but their energetic significance is profound.
The Lasso picture itself, if analyzed theoretically, would definitely need to take into account the feedback between particle and plasma dynamics, since the energy content of the particles can be large.
\nocite{1997ApJ...485..859S}
This means that a test-particle description of the collapsing trap (e.g., Somov \& Kosugi 1997; Eradat Oskoui \& Neukirch, 2014) may not capture its physics self-consistently.
\nocite{2014A&A...567A.131E}

The essential novelty of the Lasso picture is twofold: first, substantial particle acceleration must take place on closed field lines, implying that the CME shock (if that is the accelerator) must penetrate such regions; and second, that the post-event dynamics of flux bundles containing high-energy particles can facilitate their interaction with the lower atmosphere.
The magnetic retraction would help out with the number problem via the additional acceleration terms. 
In such a case the centroid position of the $\gamma$-ray source might lie well above the limb, given turbulent trapping; if loss-cone dynamics were important, the sources would be in the lower solar atmosphere.
If the CME-driven shock accelerates the particles, it must have developed in the deep atmosphere or else have flanks extending into closed-field regions not participating in the eruption.
In neither case would we expect substantial $\gamma$-ray source motion.
If SEPs could return to the Sun from well-connected open fields, somehow overcoming the mirror force (e.g. Jin \textit{et al.} 2017; Plotnikov \textit{et al.} 2017), this might not be the case.
\nocite{2017SPD....48.0303J}

\bigskip
\noindent{{\bf Acknowledgments:}} 
I am grateful for extensive discussion of these and related ideas in the ISSI team ``Energetic Ions: The Elusive Component of Solar Flares,'' chaired by Alec MacKinnon and with the participation of 
Federico Benvenuto,
Christina Cohen,
Nicole Duncan,	
Luis Fernandez Menchero, 	
S{\" a}m Krucker,	
Alec MacKinnon,	
Melissa Pesce-Rollins,
Albert Shih,	
Paulo Sim{\~ o}es,	
Rami Vainio,	
Nicole Vilmer, and	
Pietro Zucca.
Georgia De Nolfo and Jim Ryan helped by reading the first version, and I also benefited from discussions with Frederic Effenberger and Ludwig Klein.

\bibliographystyle{apj}
\bibliography{exeter}

\end{document}